\documentclass[a4paper,10pt]{article}
\usepackage[utf8]{inputenc}

\usepackage{fullpage}
\usepackage{amsmath, amsthm,amssymb}
\newtheorem{theorem}{Theorem}
\newtheorem{definition}{Definition}
\title{Computing an Evolutionary Ordering is Hard}
\author{Laurent Bulteau, Gustavo Sacomoto, Blerina Sinaimeri\\
\small Université Lyon 1; INRIA Rhône-Alpes; CNRS, UMR5558; Laboratoire de Biométrie et Biologie Evolutive
}
\date{}
\begin{document}

\maketitle
We study the problem of computing evolutionary orderings of families of sets, as introduced by Little and Campbell~\cite{Little00}.

\begin{definition}
 Let $\mathcal S$ be a family of subsets of some universe $U$. We say that $\mathcal S$ is \emph{evolutionary} if there exists an ordering of its sets $\mathcal S=\{S_1,S_2, \ldots, S_m\}$ such that:
\begin{itemize}
 \item Each set brings a new element, i.e. $S_i\nsubseteq \bigcup_{j=1}^{i-1} S_j$
 \item Each set, except the first one, has an old element, i.e. $S_i \cap \bigcup_{j=1}^{i-1} S_j \neq \emptyset$
\end{itemize}
\end{definition}

The associated algorithmic problem is the following:
\\{\sc Evolutionary Ordering}
\\{\bf Input:} A family of subsets $\mathcal S$ of some universe $U$.
\\{\bf Question:} Is $\mathcal S$ evolutionary?

We determine the computational complexity of this problem.

\begin{theorem}
 {\sc Evolutionary Ordering} is NP-hard.
\end{theorem}

By a reduction from 3-SAT. Consider a formula $\Phi$ with $n$ variables and $m$ clauses. 
Assume that each clause appears twice (i.e., $\Phi$ can be written $\Phi=\Phi'\wedge \Phi'$). 
This is not restrictive, 3-SAT is clearly hard even when restricted to this class of formulas.  For ease of presentation, assume that each literal occurs exactly $k$ times, and each clause has exactly 3 literals. Note that $2kn=3m$. 

The universe on which the sets are constructed contains the following $6n+5m$ elements:
\begin{itemize}
 \item $2n$ \emph{assignment elements}, denoted $x_i$ and $\bar x_i$ for each $1\leq i\leq n$
 \item $2n+1$ \emph{trigger elements}, denoted  $t_i$ and $\bar t_i$ for each $1\leq i\leq n$ and $\tau$
 \item $2n$ \emph{free elements}, denoted $f_i$ and $\bar f_i$ for each $1\leq i\leq n$
 \item $2kn$ \emph{literal elements}, denoted $\ell_i^h$ and $\bar \ell_i^h$ for each $1\leq i\leq n$ and $1\leq h\leq k$
 \item $2m$ \emph{clause elements}, denoted $c_j$ and $c_j'$ for each $1\leq j\leq j'$
\end{itemize}

We now create the following sets:
\begin{itemize}
 \item Two \emph{triggering sets}:
 $$T:=\{\tau\}$$
 $$T':=\{\tau, t_1,\bar t_1, t_2,\ldots, \bar t_n\}$$
 \item For each $1\leq i\leq n$, define two \emph{variable sets} and a \emph{verification set}:
 $$L_i := \{x_i, t_i, f_i, \ell_i^1, \ldots,  \ell_i^k \}$$
 $$\bar L_i := \{\bar x_i, \bar t_i, \bar f_i, \bar \ell_i^1, \ldots,  \bar \ell_i^k \}$$
 $$V_i := \{x_i, \bar x_i, c_1, c_1', c_2, c_2', \ldots c_m, c_m'\}$$ 
 \item For each $1\leq j \leq m$, where the $j$th clause uses, say, literals $\ell_1^1$, $\ell_2^1$, $\bar \ell_3^1$, deifne two \emph{clause sets}
 $$C_j := \{ \ell_1^1, \ell_2^1, \bar \ell_3^1, c_j \}$$
 $$C'_j := \{ c_j, c_j' \}$$
 
\end{itemize}

We prove that this collection of sets has an evolutionary ordering if, and only if, $\Phi$ is satisfiable.

\emph{If.} Given a truth assignment, we simply give an ordering of the sets by \emph{adding} them one by one.

\begin{itemize}

\item Start with the triggering sets $T$ and $T'$. No condition need to be satisfied for $T=\{\tau\}$, and, in $T'$, $\tau$ is old and $t_1$ is new.

\item For each variable $x_i$, add $L_i$ if $x_i$ is assigned true, $\bar L_i$ otherwise. For each one, $t_i$ (or $\bar t_i$) is old, and $f_i$ (or $\bar f_i$) is new.

\item For each clause $c_j$, add set $C_j$ followed by $C_j'$. Since the clause is satisfied, some literal $\ell_i^h$ (or $\bar \ell_i^h$) must be assigned true, 
so the corresponding element in $L_i$ (or $\bar L_i$) is old for set $C_j$. Element $c_j$ is new for set $C_j$, and then old for set $C_j'$.  Element $c_j'$ is new for set $C_j'$.

\item For each variable $x_i$, add the verification set $V_i$. Element $c_1$ is old. If $x_i$ is assigned true (resp. false), then element $\bar x_i$ (resp. $x_i$) is new.

\item For each variable $x_i$, add $\bar L_i$ if $x_i$ is assigned true, $L_i$ otherwise. For each one, $t_i$ (or $\bar t_i$) is old, and $f_i$ (or $\bar f_i$) is new.

\end{itemize}
Overall, we have an ordering of the sets where each one has an old and a new element: the set is evolutionary.

\emph{Only if.} Assume that our family of sets is evolutionary and consider such an ordering.
Note that $T=\{\tau\}$ must be the very first set of this ordering (since otherwise iot cannot contain both old and new elements). This means that all other sets have an old and a new element.
Write $\mathcal A$ for the family of the sets $L_i$ and $\bar L_i$ that appear before their corresponding verification sets $V_i$. 
We make the following observations.

First, for each clause $c_j$ and each variable $x_i$, set $C'_j$ appears before  $V_i$. This is because $C'_j\subset V_i$.

Now, for each variable $x_i$, it is not possible to have both $L_i\in \mathcal A$ and $\bar L_i\in \mathcal A$. 
Otherwise, $V_i$ would not have any new element, since $V_i\subseteq L_i \cup \bar L_i \cup \bigcup_{j=1}^m C'_j$ and each $C'_j$ is already before $V_i$. 
Thus, we design a truth assignment such that $x_i$ is true if $L_i\in \mathcal A$, and false otherwise.
This way, for each $L_i$ or $\bar L_i$ in $\mathcal A$, the corresponding literal ($x_i$ or $\bar x_i$), is assigned true.

It remains to show that the assignment satisfies formula $\Phi$. 
Consider each clause $c_j$. 
First, $C_j$ appears before  $C'_j$.  
Indeed, the only sets intersecting $C'_j$ are $C_j$ and each $V_i$. 
Remember that $V_i$s appear after $C'_j$, so the old element in $C'_j$ must be from $C_j$, and $C_j'$ appears after $C_j$.
It follows that the old element of $C_j$ cannot be $c_j$, hence it is a literal element $\ell_i^h$ or $\bar \ell_i^h$. 
So the corresponding variable set $L_i$ or $\bar L_i$ must be before $C_j$, hence before $C'_j$ and $V_i$. 
Overall, for each clause,  one of $L_i$ or $\bar L_i$ corresponding to a literal of the clause is in $\mathcal A$, and the literal is satisfied by our assignment. 
So the whole formula $\Phi$ is satisfiable.

\end{document}